\documentclass[a4paper,fleqn,usenatbib,letters]{mnras}
\usepackage{graphicx}
\usepackage{amssymb}
\usepackage{amsmath}

\newcommand{\mbfvD}{\mathbf{v}_{tot}}
\newcommand{\mbfvR}{\mathbf{\Theta}}
\title[3D Orientation of Compact HVCs]{Three-Dimensional Orientation of Compact High Velocity Clouds} 
\author[Heitsch et al.]
{\parbox{\textwidth}{F. Heitsch,$^{1}$\thanks{E-mail: fheitsch@unc.edu}
B. Bartell,$^{1}$
S.E. Clark,$^{1,2}$
J.E.G. Peek,$^{2,3}$
D. Cheng$^{1}$ and M. Putman$^{2}$}\vspace{0.4cm}\\
\parbox{\textwidth}{
$^{1}$Department of Physics and Astronomy, University of North Carolina Chapel Hill, Chapel Hill, NC 27599-3255, U.S.A\\
$^{2}$Department of Astronomy, Columbia University, 550 W 120th St, New York, NY 10027, U.S.A\\
$^{3}$Space Telescope Science Institute, 3700 San Martin Dr, Baltimore, MD 21218, U.S.A
}}

\date{Accepted 2016 June 21 Received 2016 June 21; in original form 2016 March 24}

\pubyear{2016}

\begin{document}
\label{firstpage}
\pagerange{\pageref{firstpage}--\pageref{lastpage}}
\maketitle

\begin{abstract}
We present a proof-of-concept study of a method to estimate the inclination angle of compact high velocity clouds (CHVCs), 
i.e. the angle between a CHVC's trajectory and the line-of-sight. The inclination angle is derived from the CHVC's morphology 
and kinematics. We calibrate the method with numerical simulations, and we apply it to a sample of CHVCs drawn from HIPASS. 
Implications for CHVC distances are discussed.
\end{abstract}



\section{Motivation}
The Galactic halo hosts a population of neutral hydrogen clouds whose line-of-sight velocities are inconsistent
with Galactic rotation \citep{1997ARA&A..35..217W}. These High Velocity Clouds (HVCs) range from large ``complexes''
of many degrees to structures at the resolution limit. Their diversity suggests different
origins \citep{1997ARA&A..35..217W,2012ARA&A..50..491P}.
Distances to HVCs are key to the origin question.
The most accurate constraints stem from absorption line studies
\citep{2001ApJS..136..463W,2007ApJ...670L.113W,2006ApJ...638L..97T,2008ApJ...684..364T,2015A&A...584L...6R}, 
yet these are only available for structures of large angular extent, and therefore are biased to near objects.
Indirect distances via H$\alpha$ emission use the UV flux escaping from the disk and ionizing the HVCs 
\citep{2003ApJ...597..948P}. Uncertainties arise from determining the escape fraction of ionizing UV photons, though the 
patchiness of the disk interstellar gas ceases to be of concern for $|z|>10$~kpc \citep{2001ASPC..240..369B,2007ApJ...656..907P}.
\citet{2008A&A...485..457O} uses a putative origin to constrain distances of CHVCs spatially associated with the 
Magellanic complexes \citep[also][]{2008ApJ...674..227P,2012ApJ...758...44S}. Distance constraints based on cloud 
kinematics assume a terminal velocity for HVCs \citep{1997ApJ...481..764B} or rely on differential drag due to the interaction 
with the background medium \citep{2007ApJ...656..907P}. Both of these methods require the inclination angle between
the cloud's trajectory and the line-of-sight. Full trajectory information has
been inferred in only a few cases 
(Smith Cloud: \citealp{2008ApJ...679L..21L,2016ApJ...816L..11F}; Complex GCN: \citealp{2010MNRAS.408L..85J}).

We will focus our attention on compact high velocity clouds (CHVCs), many of whom show a head-tail structure, consisting of a 
cold, dense core, and a more diffuse, warmer tail \citep{2000A&A...357..120B,2001A&A...370L..26B}. This morphology suggests 
that CHVCs interact with the ambient medium during their passage through the Galactic halo 
\citep{2000A&A...357..120B,2006ApJ...653.1210S,2007ApJ...656..907P,2011MNRAS.418.1575P}. Because of their small angular extent, 
distance estimates to CHVCs stem mostly from assumed association with larger complexes 
\citep{2008ApJ...674..227P,2011MNRAS.418.1575P}. Yet, an independent method is desirable. Because of their interaction with 
the ambient gas, CHVCs could in principle be used to gain information about the elusive gaseous component of the Galactic 
halo \citep{2007ApJ...656..907P}.

Instead of aiming directly at getting distances to CHVCs, we propose a method to determine the three-dimensional 
orientation of CHVCs and thus their full, three-dimensional velocity $\mbfvD$. Consequences for distance constraints are 
discussed in Sec.~\ref{ss:distancebracket}.


\section{The Method}\label{s:method}
\begin{figure}
\begin{center}
\includegraphics[width=0.8\columnwidth]{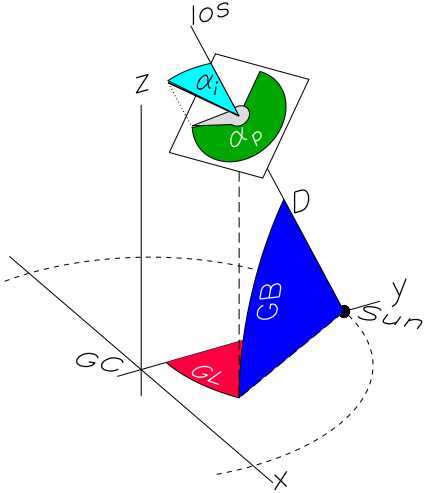}
\end{center}
\caption{\label{f:coords} Definition of the coordinate system for CHVC orientation. 
         The position angle $\alpha_p$ (green) and the inclination angle $\alpha_i$ (cyan) are defined in the local 
         (GL,GB) patch, at an (unknown) distance $D$ from the Sun. A cartoon CHVC mapped within the local (GL,GB) patch is outlined
         in grey.} 
\end{figure}

The coordinate system is set by the local $(GL,GB)$ patch describing the plane-of-sky, and by the (unknown) cloud distance 
$D$ along the line-of-sight (Fig.~\ref{f:coords}).  The three-dimensional orientation of a CHVC requires two angles:
The inclination angle $0\leq\alpha_i\leq\pi$ describes the angle between the cloud {\em tail} and the line-of-sight, 
with the tail pointing away from the observer for $\alpha_i=0$. The position angle $0\leq \alpha_p < 2\pi$ is 
counted counter-clockwise starting with $\alpha_p=0$ for the cloud's tail pointing toward Galactic North.
The local coordinate patch is assumed to be rectangular -- hence the limitation to CHVCs. 

\begin{figure}
  \includegraphics[width=\columnwidth]{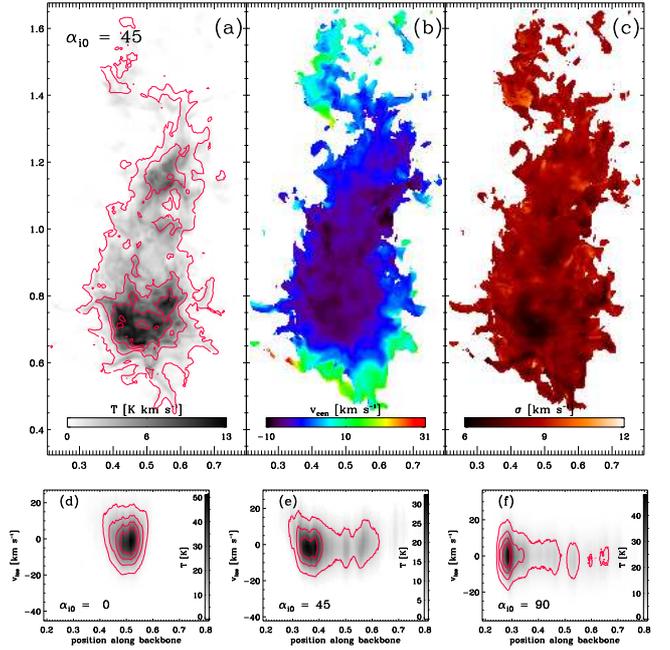}
  \caption{\label{f:modelmaps} (a) Integrated intensity, (b) centroid velocity,  and (c) velocity dispersion of a model CHVC
  (model Wb1a15b of \citet{2009ApJ...698.1485H}), traveling at $45^{\circ}$ to the observer. Contours are given at 
  $[1,5,9,13]$~K~km~s$^{-1}$. (d,e,f) Position-velocity plots for inclination angles $\alpha_{i0}=0,45,90^{\circ}$. Contours 
  correspond to $[5,15,25,35]$~K.}
\end{figure}

The goal is to relate the cloud shape in position-velocity space to the inclination angle. We demonstrate the process with
the help of a simulation (Fig.~\ref{f:modelmaps}) of a CHVC traveling at $\alpha_i=45^{\circ}$ toward the observer.  
The simulation (model Wb1a15b of \citet{2009ApJ...698.1485H}, see their table~1 and fig.~2) 
is a wind-tunnel experiment, in which an 
initially spherical cloud of (in this case) radius $50$~pc and density $0.1$~cm$^{-3}$ is exposed to a wind of $150$~km~s$^{-1}$ 
and a density of $10^{-5}$~cm$^{-3}$. The simulation generated $\sim 30$ 3D data sets consisting of gas density, velocity and 
temperature. These are converted into position-position-velocity cubes by selecting for gas with a temperature of $T<10^4$~K 
(assumed to be neutral hydrogen), rotating by the desired inclination angle $\alpha_{i0}$, and then calculating channel maps 
with $\Delta v=1$~km~s$^{-1}$ assuming optically thin HI-21~cm emission. Peak column densities reach 
$\sim 3\times 10^{19}$~cm$^{-2}$. These channel maps are then used for further analysis. 

The position angle $\alpha_p$ is determined by
fitting ellipses to the integrated intensity maps. Since the orientation of the ellipse is
degenerate with $\pi$, we identify the tail of the cloud as the direction in which the cloud extends farthest from the
column density peak (i.e. the location of the head). This assumes that the clouds have a head-tail structure.
To estimate the inclination angle, we define the cloud's "backbone" (i.e. the line through the cloud's center-of-mass at the 
determined $\alpha_p$), along which spectra are taken to construct a position-velocity map (Fig.~\ref{f:modelmaps}e-g). 
For a CHVC moving toward the observer, the (dense) core will appear at more negative velocities and the tail at more positive 
ones, hence the CHVC will be asymmetric along the velocity axis. Yet, along the position axis, the CHVC will appear more or 
less symmetric (Fig.~\ref{f:modelmaps}d). If the CHVC moves perpendicularly to the line-of-sight, head and tail can be clearly 
identified, resulting in an asymmetry in position. Yet, the CHVC will appear symmetric in velocity space, since the gradient 
along the cloud backbone due to the differential drag will not be discernible, and only thermal and turbulent motions within 
the CHVC will contribute to the velocity signature (Fig.~\ref{f:modelmaps}f). A CHVC traveling at e.g. $45^{\circ}$ 
to the observer will appear asymmetric both in position and velocity (Fig.~\ref{f:modelmaps}e).

We calculate the observable asymmetry of the CHVC's gas distribution with respect to its center-of-mass.  The asymmetry in
position is given by
\begin{equation}
  a_p\equiv\frac{\Delta_{2p}-\Delta_{1p}}{\Delta_{1p}+\Delta_{2p}},\label{e:asymmetry}
\end{equation}
with $-1\leq a_{p} \leq 1$. The one-sided dispersions $\Delta_{1p,2p}$ refer to the 
CHVC extent to lower/higher values in position with respect to the center-of-mass, e.g.
\begin{equation}
  \Delta_{1p} = \left(\sum_{p<p_c,v}(p-p_c)^2\,T(p,v)\bigg/\sum_{p<p_c,v}T(p,v)\right)^{1/2},
\end{equation}
where $T(p,v)$ is the position-velocity map, $p_c$ is the center-of-mass position, and the summation extends over all $p<p_c$ 
(along the horizontal axis in Fig.~\ref{f:modelmaps}d-f), and over the whole velocity range. 
For $\Delta_{2p}$, the summation extends over $p>p_c$. The velocity extents $\Delta_{1v,2v}$ are constructed similarly, along the
vertical (velocity) axis of the position-velocity plot. Other measures
of cloud extent, such as $50$\% contours, give similar results. Since the accuracy of $\Delta_{1,2}$ depends on the map resolution,
spatial and velocity resolution of the telescope will affect the result. 
A CHVC with the tail pointing
toward positive $p$ has $a_p>0$, and a CHVC moving toward the observer (tail toward positive $v$) has $a_v>0$.
Since the asymmetries are normalized, and if we assume (to first order) a linear relationship between the velocity
and the position along the cloud's tail \citep[see e.g.][]{2001A&A...370L..26B}, we can calculate the inclination angle as
\begin{equation}
  \alpha_i = \arctan\left(\frac{a_p}{a_v}\right).\label{e:pitchangle}
\end{equation}

\begin{figure*}
  \includegraphics[width=\textwidth]{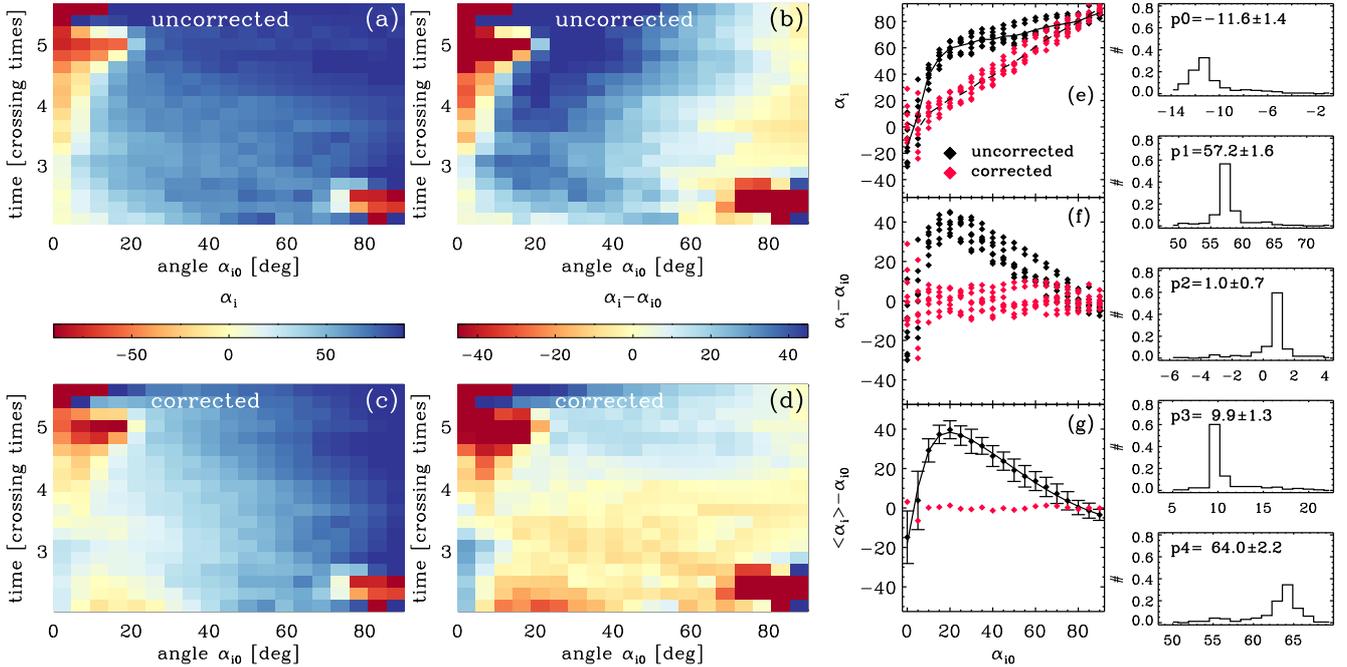}
  \caption{\label{f:alphatests}(a) Colour map of the uncorrected inclination angle estimate (equation~\ref{e:pitchangle}) 
          depending on the rotation angle $\alpha_{i0}$, and on
          the model CHVC evolution time. Large ``secular'' differences occur when
          substantial fragments are stripped off the CHVC. 
          (b) Uncorrected inclination angle residuals scaled between $\pm 45^{\circ}$. (c) Corrected inclination angle estimates,
          and (d) corrected residuals. (e) Uncorrected (black) and corrected (red) inclination angles, and (f) their residuals. 
          (g) Residuals calculated from averaged uncorrected inclination angle (black), empirical fit 
          (line, see equation~\ref{e:fitres}), and resulting corrected residuals
          (red). The right column gives the fit parameter distributions and values.}
\end{figure*}

To test the method, position-position-velocity cubes are generated for the CHVC model of Fig.~\ref{f:modelmaps}, for a series of rotation
angles $\alpha_{i0}$.
"Spectra" (position-velocity plots) are taken along the long axis of the cloud (Fig.~\ref{f:modelmaps}d-f), from which we derive the
inclination angle estimate $\alpha_i$. Fig.~\ref{f:alphatests} summarizes the reliability of the inclination angle estimates.
Panels (a) and (b) show  $\alpha_i$ as derived from 
equation~\ref{e:pitchangle}, and its residuals $\alpha_i-\alpha_{i0}$.
Apart from occasional large deviations due to substantial fractions of gas 
being stripped off the CHVC, the residuals depend systematically on the model rotation angle $\alpha_{i0}$ (Fig.~\ref{f:alphatests}f). 
Therefore, we attempt to improve on equation~\ref{e:pitchangle} by fitting a heuristic function to the residuals; we average
over the cloud evolution time (Fig.~\ref{f:alphatests}g, the error bars are errors on the mean). The resulting
corrected values are shown in red in Fig.~\ref{f:alphatests}e through \ref{f:alphatests}g, and Fig.~\ref{f:alphatests}b,d. 
The fitting function is given by
\begin{equation}
  f(\alpha_{i0}) = p_0\,\tanh\frac{\alpha_{i0}-p_2}{p_3}\,\exp\frac{\alpha_{i0}-p_2}{p_4}+p_1.\label{e:fitres}
\end{equation}
Parameter distributions and values derived from the Metropolis-Hastings algorithm used to fit equation~\ref{e:fitres} 
are given in the right column of Fig.~\ref{f:alphatests}.

\section{Application to CHVCs}\label{s:application}
\begin{figure*}
  \includegraphics[width=\textwidth]{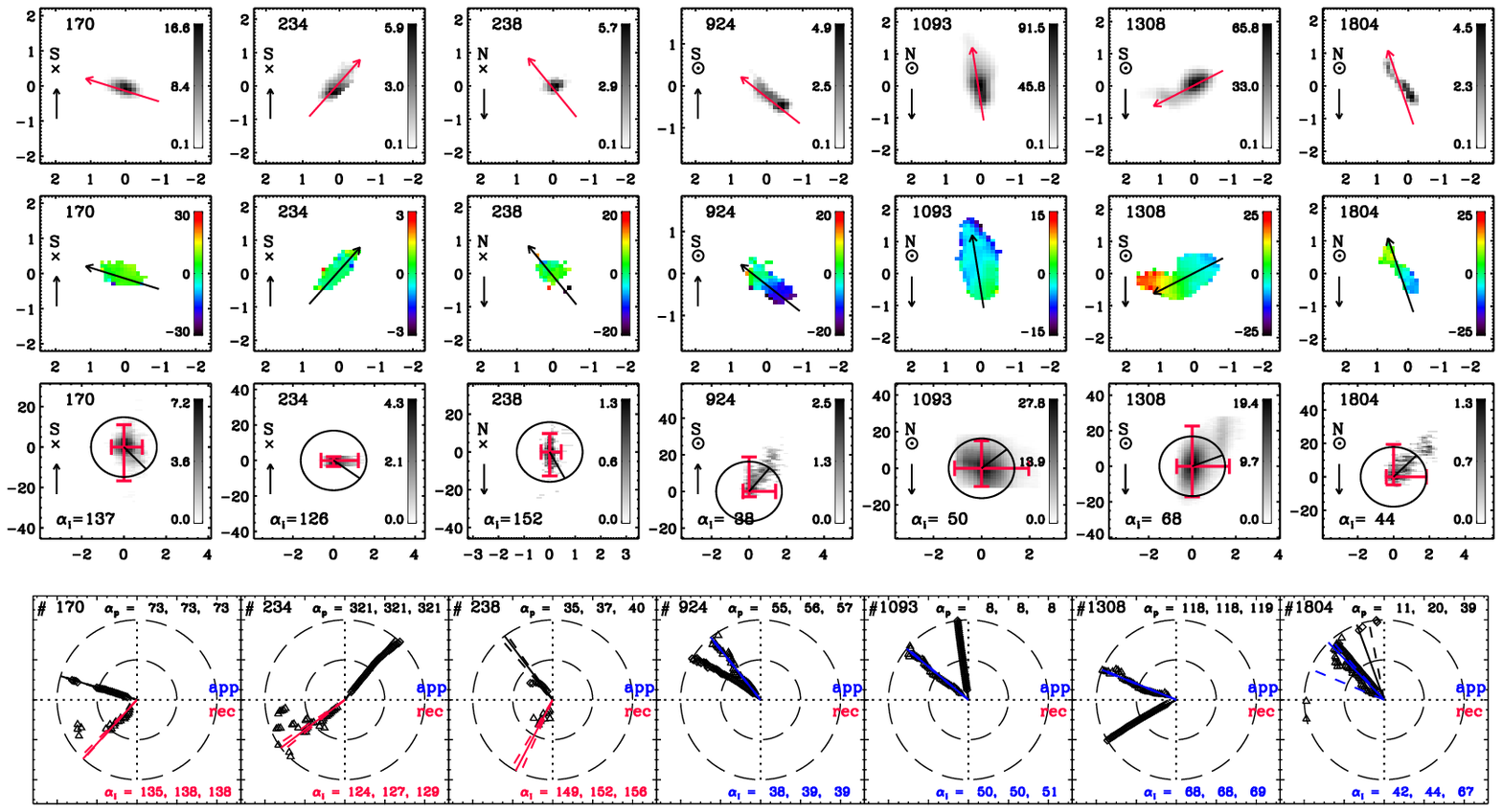} 
  \caption{\label{f:cloudmaps}Top to bottom: Integrated intensity for CHVCs selected from HIPASS 
  \citep{2002AJ....123..873P}, velocity centroid maps, position-velocity plots, and derived position and inclination angles. 
  Red arrows denote the direction of the cloud tail. Positions in $(GL,GB)$ are relative to the cloud's catalogued coordinates. 
  Letters $N$ and $S$ indicate Northern or Southern Galactic hemisphere. Black vertical arrows indicate whether the cloud is 
  moving toward lower (downward) or higher $|z|$ (upward). Clouds approaching the observer ($\alpha_i < 90^\circ$) are denoted
  by a dotted circle, otherwise by a ``x''. The position axis in the pv-plots is counted from head to tail. 
  {\em Bottom row:} Position angle $\alpha_p$ (black symbols, lines) and inclination angle $\alpha_i$ (blue or red symbols, lines).
  Both angles $\alpha_p$ and $\alpha_i$ are counted counter-clockwise from the top, with $0\leq\alpha_p\leq360$ and $0\leq\alpha_i\leq 180$. 
  Long dashed circles indicate $S/N$ values of $20$ and $40$. Solid lines refer to the median value, short dashed lines to the 
  lower and upper quartile. These values are given also at the top and bottom of each panel (lower quartile, median,
  upper quartile). The horizontal dashed line separates inclination angles for approaching (blue) and receding (red) CHVCs.}
\end{figure*}

We apply the inclination angle estimate to selected CHVCs drawn from HIPASS \citep{2002AJ....123..873P}.
We select with a slight preference for head-tail clouds, yet we note that the head-tail structure would not show
when the CHVC is traveling along the line-of-sight.
The top two rows of Fig.~\ref{f:cloudmaps}
show the integrated intensity and centroid velocity. HIPASS catalogue numbers 
are given in each panel. We apply a selection ellipse around 
the CHVC structure of interest, removing unassociated emission, both in $(GL,GB)$-space and in 
$v_{lsr}$-space.

We determine the angles $\alpha_p$ and $\alpha_i$ for a sequence of increasing signal-to-noise 
values ($S/N=[5,40]$ in steps of $1$). For $S/N<5$,
angle estimates were generally unreliable in our sample. The bottom row 
of Fig.~\ref{f:cloudmaps} summarises the derived angles for the selected CHVCs. Shown are the median values (solid lines)
including lower and upper quartiles (dashed lines), to highlight the uncertainties in the angle estimates. The position angle
$\alpha_p$ can be determined within $< \pm 3^\circ$ (exception: cloud 1804, whose position angle ``drifts'' with $S/N$). For
well-defined clouds, the inclination angles show similar ranges. To further assess the reliability of the angle estimates, we calculate
$\alpha_i$ for all $0\leq \alpha_p\leq 360$ and for all $S/N$. In the resulting map of $\alpha_i(\alpha_p,S/N)$
we search for ``consistent'' $\alpha_i$ values, i.e. for regions in $(\alpha_p,S/N)$ space across which $\alpha_i$ does not change by 
more than $2^\circ$. These regions are usually extended over a large range in $S/N$, while for inconsistent solutions, $\alpha_i$ 
varies strongly with $\alpha_p$. The largest of these regions is taken as the solution. The resulting angle estimates are consistent 
with the direct fits described above.

\section{Discussion}\label{s:discussion}

\subsection{A Method to Constrain CHVC Distances}\label{ss:distancebracket}
We explore whether the full cloud orientation can be used to derive distance constraints of CHVCs via the velocity of a CHVC 
relative to its background medium, $v_{rel}$. This requires several assumptions. It is not our intent that these 
be necessarily correct, but that they are sufficiently plausible to outline the method.
The goal is to calculate $v_{rel} = |\mbfvD-\mbfvR|$ along the line-of-sight at a given $(GL,GB)$ for a range of distances $D$. 
Here, $\mbfvD$ is the three-dimensional velocity of the CHVC in Galactic cartesian coordinates, and $\mbfvR$ is
the three-dimensional (halo) rotation velocity of the background medium. All velocities are relative to
the Galactic Standard of Rest (GSR).
Since $\mbfvD$ is constant along the line-of-sight, but $\mbfvR$ will change with $D$, $v_{rel}=v_{rel}(D)$. If we have additional
information on $v_{rel}$, such as a terminal velocity $v_{max}$ at which CHVCs can move with respect to the background medium, distances 
$D$ can be identified for which $v_{rel}\leq v_{max}$.

Setting $\mbfvR$ requires a Galactic halo rotation model.
For demonstration, we combine the rotation curve model of \citet{1989ApJ...342..272F} with an exponential drop-off in $z$, 
reproducing the linear gradient of $-22$~km~s$^{-1}$ derived by \citet{2008ApJ...679.1288L}. Our halo rotation model then reads as
\begin{equation}
  |\mbfvR|\equiv v_R(R_{xy},z) = (109+108 R_{xy}^{0.0042})\,e^{-|z|/10},\label{e:vrot}
\end{equation}
with $R_{xy}$ and $z$ in kpc. 
The radius $R_{xy}$ gives the Galactocentric distance in the plane, with the full Galactocentric radius being 
$R=(R^2_{xy}+z^2)^{1/2}$. 

There are several options to constrain $v_{rel}$, such as setting $v_{max}$ to the terminal velocity due to hydrodynamical drag 
\citep{1997ApJ...481..764B}, estimating $v_{rel}$ based on differential drag analysis of the CHVC \citep{2007ApJ...656..907P}, or 
limiting $v_{max}$ to the sound speed of the background medium for sufficiently diffuse CHVCs. Based on our models \citep{2009ApJ...698.1485H},
we choose the latter and set $v_{max}=c_s=100$~km~s$^{-1}$. Other options
will be explored in a future contribution.

Table~\ref{t:cloudparam} summarises the estimated parameters for the seven CHVCs shown in Fig.~\ref{f:cloudmaps} together with 
a few other CHVCs selected from HIPASS. Roughly $50$\% of the sample CHVCs have near distance constraints 
(at $c_s=100$~km~s$^{-1}$). Most remaining CHVCs show relative velocities $v_{rel}>200$~km~s$^{-1}$, and thus do not lead to a 
distance constraint. The value of $|\mbfvD|$ depends strongly on $\alpha_i$: At $\alpha_i=90,270^{\circ}$, $|\mbfvD|$ 
cannot be reconstructed.

Though none of the observed CHVCs have (previous) direct distance constraints, 
many of them are potentially related to larger HVC complexes with constraints from their position-velocity 
proximity \citep{2008ApJ...674..227P,2011MNRAS.418.1575P}.  In the Southern sky, the majority of the HVCs (and the CHVCs 
in Table~\ref{t:cloudparam}) can be associated with the Magellanic System and though the distance to the Magellanic 
complexes are unknown, the Magellanic Clouds themselves are at $50$-$60$~kpc and the associated clouds are 
expected to be further away than the lower distance limits $D_{lo}$ in Table~\ref{t:cloudparam}.
One strong comparison point in our sample is cloud 238, which is in the position-velocity vicinity of the 
tail of Complex C. Complex C has a direct distance constraint of $10$~kpc \citep{2008ApJ...684..364T}, and we 
find this cloud has a lower distance estimate of $10$~kpc.  Within the uncertainties, the 
results of the method are thus far consistent with 
existing distance constraints.

\begin{table}
\caption{Selected HIPASS CHVC parameters.
[1] HIPASS number. [2] Galactic longitude GL. [3] Galactic latitude GB. [4] Position angle $\alpha_p$.
[5] Inclination angle $\alpha_i$. [6] Lower distance limit $D_{lo}$. [7] Total velocity $|\mbfvD|$. [8] Cloud approaching
($\odot$) or receding ($\otimes$), and moving toward ($\downarrow$) or away from ($\uparrow$) disk. [9] Possible
association with known HVC complexes. Complexes in square brackets do not have distance constraints. For identification of
the HVC complexes, see \citet{2006A&A...455..481K,2012ARA&A..50..491P}.}
\label{t:cloudparam}
\begin{tabular}{*{9}{c}}
\hline
[1] & [2] & [3] & [4] & [5] & [6] & [7] & [8] & [9] \\
\hline
 170&$ 16.8$&$-25.0$&$ 73$&$138$&    $-$&$240$&$\otimes  \uparrow$&GCN\\
 234&$ 24.5$&$ -1.8$&$321$&$127$&    $-$&$324$&$\otimes  \uparrow$&GCN\\
 238&$ 24.8$&$  8.8$&$ 37$&$152$&$ 10.0$&$ 24$&$\otimes\downarrow$&C\\
 924&$258.5$&$-39.1$&$ 56$&$ 39$&$ 21.7$&$ 85$&  $\odot  \uparrow$&LA\\
1093&$271.0$&$ 10.8$&$  8$&$ 50$&$ 20.9$&$ 72$&  $\odot\downarrow$&LA\\
1308&$285.0$&$-16.1$&$118$&$ 68$&$ 22.4$&$  2$&  $\odot\downarrow$&LA\\
1804&$334.8$&$ 30.7$&$ 20$&$ 44$&    $-$&$234$&  $\odot\downarrow$&L\\
  48&$  3.9$&$-63.7$&$ 70$&$124$&    $-$&$270$&$\otimes  \uparrow$&MS?\\
 200&$ 21.2$&$-61.2$&$236$&$ 24$&$ 12.3$&$ 59$&  $\odot\downarrow$&MS?\\
 632&$224.1$&$-17.0$&$  8$&$139$&$ 24.0$&$ 81$&$\otimes  \uparrow$&MS?\\
 648&$226.6$&$-33.4$&$ 90$&$ 25$&$  7.7$&$ 27$&  $\odot  \uparrow$&MS?\\
1221&$279.1$&$-16.7$&$211$&$ 49$&$  8.4$&$ 97$&  $\odot\downarrow$&LA\\
1616&$316.9$&$-76.8$&$ 94$&$ 67$&    $-$&$166$&  $\odot\downarrow$&MS\\
1806&$335.0$&$ 16.1$&$191$&$ 50$&$ 29.4$&$ 41$&  $\odot  \uparrow$&[WD]\\
\hline
\end{tabular}
\end{table}

The weakest link in these distance constraints is the choice of a halo rotation model. Increasing 
the characteristic scale from $10$ to $20$~kpc in equation~\ref{e:vrot} (and thus flattening the drop-off of
$\mbfvR$ with $z$) increases all the distance constraints by a factor of $\sim2$. Halo rotation models without $z$-dependence
\citep{2016ApJ...822...21H} do not yield results {\em if we assume} $v_{rel}\lesssim100$~km~s$^{-1}$.   
We interpret this as a limitation of our assumptions regarding $v_{rel}$ rather than a limitation of the method itself.

\subsection{Caveats}\label{ss:caveats}
\noindent{\em Residual Fitting} 
Correcting the inclination angle estimate (equation~\ref{e:pitchangle}) by fitting the residuals raises the question about the
physical motivation for equation~\ref{e:fitres}. Equation~\ref{e:pitchangle} assumes that the velocity gradient
along the tail, caused by deceleration of the cloud gas, is linear. This is not necessarily correct 
\citep[][see also Fig.~\ref{f:modelmaps}]{2001A&A...370L..26B,2007ApJ...656..907P};  material directly behind the cloud is
expected to travel nearly at the same velocity as the cloud. Velocities
close to the cloud speed reduce the velocity asymmetry $a_v$, thus overestimating $\alpha_i$.
The fit parameters might also depend on environmental factors, such as the ambient density, and the absolute cloud velocity. 
These dependencies and their quantification can only be explored with a larger model grid, which is beyond the scope 
of this paper.

\noindent{\em Effect of Background Flow on $\alpha_i$}
Since the CHVCs in our sample are identified via HI emission, their $\alpha_i$ estimates rest on the assumption that
the neutral gas interacts directly with the background halo. Yet, there is evidence for substantial ionized envelopes co-moving with
HVCs \citep{2009ApJ...703.1832H,2009ApJ...702..940L,2012MNRAS.424.2896L}. For 
a CHVC moving at a velocity $|\mbfvD|=v_{rel}+v_{env}$ with respect to an ionized envelope traveling in the same direction at 
$v_{env}$, the reduced drag results in a smaller spread along the velocity axis in the position-velocity plot, and therefore in 
a pitch angle biased toward $90^{\circ}$. This in turn increases the inferred total velocity $\mbfvD$. On the 
other hand, the observed radial velocity combines the line-of-sight component of the HI CHVC and the ionized envelope. 
Therefore, our method tends to {\em overestimate} $|\mbfvD|$ if the CHVC is moving within a larger ionized envelope. Yet, if 
the line-of-sight component of the envelope's velocity -- and therefore the line-of-sight component of $v_{rel}$ -- is known, 
the velocity spread in the position-velocity plot correctly refers to $v_{rel}$, and thus $\alpha_i$ is not affected by the 
ionized envelope.

\noindent{\em Effects of Cloud Evolution on $\alpha_i$}
The interaction of the CHVC with the ambient gas leads to turbulent 
structures, and occasionally to large ``chunks'' of the cloud being ripped off. Such ``secular'' events can affect the estimates 
for $\alpha_i$ and $\alpha_p$. The strong time variations in the residuals of $\alpha_i$ (Fig.~\ref{f:alphatests}b,d) 
are caused by this effect. 

The $\alpha_i$ estimate relies on the translation of the effect of the hydrodynamic drag on the CHVC's tail
into centroid velocity profiles. The method assumes a monotonic centroid velocity profile, i.e. for a 
cloud moving at an angle toward the observer, the head would have the most negative velocities, and the tail the most positive 
ones. Yet, the centroid velocity map of Fig.~\ref{f:modelmaps} demonstrates that this need not be the case
\citep[also][]{2001A&A...370L..26B}. The swath of ``green'' (less negative) velocities at the head of the cloud is caused 
by material flowing around the cloud away from the observer. 


\section{Summary}\label{s:summary}
We present a method to determine the three-dimensional orientation of CHVCs. The inclination angle is derived from
asymmetries in the intensity distribution of a CHVC's position-velocity plot (Figs.~\ref{f:modelmaps} and \ref{f:cloudmaps}).
We test the method with the help of numerical simulations of CHVCs and identify possible systematic effects on the inclination
angle estimate. When applied to CHVCs drawn from HIPASS, the method is returning results that are stable with increasing 
signal-to-noise. The method can be improved by a more detailed analysis of the position-velocity plots, and by a more rigorous 
statistical treatment. Applications to clouds being ablated in other astrophysical environments seem obvious.

We discuss the possibility to constrain distances by assuming a limiting CHVC velocity with respect to the background medium.
Such a limit constrains the possible locations of a CHVC along its line-of-sight, given a Galactic halo rotation model.
We find lower distance limits for $\sim 50$\% of the selected HIPASS CHVC sample. The estimates are consistent with previous
distance constraints.


\section*{Acknowledgements}
We thank the referee for a very thorough and concise report.
This work was partially supported by UNC Chapel Hill, and it has made use of NASA's Astrophysics Data System.

\bibliographystyle{mnras}
\bibliography{./references}


\bsp
\label{lastpage}
\end{document}